\documentclass[twocolumn,showpacs,amsmath,amssymb]{revtex4}
\usepackage{dcolumn}
\usepackage{bm}
\usepackage{graphicx}
\usepackage{color}

\begin{document}

\title{Conductance quantization and snake states in graphene
magnetic waveguides}
\author{T.~K. Ghosh, A. De Martino, W. H\"ausler, L. Dell'Anna, and R. Egger}
\affiliation{
Institut f\"ur Theoretische Physik, Heinrich-Heine-Universit\"at,
D-40225  D\"usseldorf, Germany}
\date{\today}

\begin{abstract}
We consider electron waveguides (quantum wires) in graphene 
created by suitable inhomogeneous magnetic fields. 
The properties of uni-directional snake states 
are discussed.  For a certain magnetic field profile, two 
spatially separated counter-propagating snake states are formed, leading
to conductance quantization insensitive to backscattering by
impurities or irregularities of the magnetic field.
\end{abstract}

\maketitle

The physics of monolayer graphene devices has recently 
attracted a great deal of attention \cite{exp,review}.
>From a fundamental perspective, one can hope to relate
experimental observations to the mathematical properties
of two-dimensional massless Dirac-Weyl quasiparticles.
The pseudo-relativistic dispersion relation with 
Fermi velocity $v_F\approx 10^6$~m/sec is intimately connected to
the sublattice structure: the basis of the graphene
honeycomb lattice contains two carbon atoms, giving
rise to an isospin degree of freedom.  Graphene has also been
suggested as new material system for device applications \cite{review}.   
In this paper, we pose (and affirmatively answer) the 
question whether quantum wires with 
quantized conductance can be formed in graphene.
Such electron waveguides are indispensable parts of 
any conceivable all-graphene device. In lithographically
formed graphene `ribbons', the electronic bandstructure is
theoretically expected to very sensitively depend on the width and on
details of the boundary \cite{ribbonthe}.
On top of that, disorder and structural inhomogeneity 
are substantial in real graphene \cite{irreg}.
For narrow graphene ribbons or electrostatically formed
graphene wires \cite{milton}, conventional
conductance quantization thus seems unlikely \cite{kats}.
This expectation is in accordance with
recent experiments \cite{ribbon}.  

\begin{figure}
\includegraphics[width=0.4\textwidth]{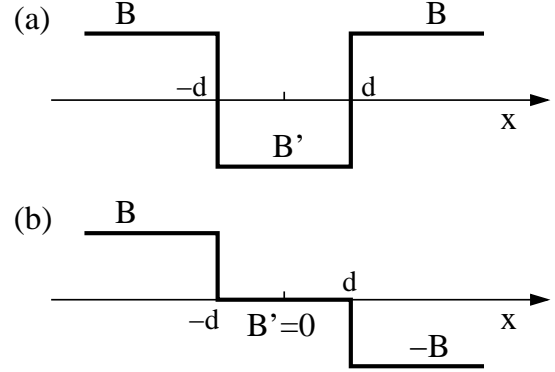}
\caption{ Magnetic field profile \eqref{field} for magnetic 
waveguide (homogeneous along $y$ direction). (a)
Case $\sigma=1$. For $B'<0$, counter-propagating  pairs of
snake states are possible.   (b) Case $\sigma=-1$, with uni-directional
propagating snake states.  
 \label{fig1} }
\end{figure}

Contrary to such pessimism, we here demonstrate that
by designing a suitable {\sl inhomogeneous} magnetic field, 
a magnetic waveguide 
can be built that indeed allows for the perfectly
quantized two-terminal conductance $4e^2/h$ (including spin and valley
degeneracy) even when disorder is present.  The disorder insensitivity is 
based on a spatial separation of the left- and right-moving `snake'
states found under the model geometry 
shown in Fig.~\ref{fig1}(a). This is reminiscent of the edge states encountered
 in the integer quantum Hall regime \cite{chang}, but here refers to a 
completely different microscopic picture.  Such {\sl double-snake states}
develop in the regime $B>0$ but $B'<0$, while an individual snake
state is uni-directional and already found in the setup of Fig.~\ref{fig1}(b).
Magnetic barrier technology is well developed \cite{snakeexp,expbar,lee}
and its application to graphene samples appears to pose no 
fundamental problems \cite{private}.  In fact, snake states were experimentally
studied in other materials \cite{snakeexp,rect}, 
mainly motivated by the quest for electrical rectification. 
 On the theory side, for Schr\"odinger fermions, 
the magnetic field profile in Fig.~\ref{fig1}(a) 
(but only for $B'=0$) was discussed
in Ref.~\cite{peeters1}, and asymmetric cases as
in Fig.~\ref{fig1}(b) were studied by a number of authors \cite{snake}.
For the Dirac-Weyl quasiparticles encountered in graphene,
however, such calculations were not reported.
Inhomogeneous magnetic fields in graphene  were
discussed by several of us \cite{prl}, and we employ
that framework in our proposal of magnetic waveguides in graphene.

For a static orbital magnetic field with perpendicular component $B(x,y)$,
the time-independent Dirac-Weyl equation for the 
quasiparticle isospinor $\Psi(x,y)$
at energy $E=v_F \epsilon$ reads (we put $\hbar=1$)
\begin{equation}\label{h0}
\mbox{\boldmath{$\sigma$}} \cdot \left(-i \mbox{\boldmath{$\nabla$}} +
\frac{e}{c} {\bf A}\right) \Psi= \epsilon \Psi,
\end{equation}
where following Ref.~\cite{prl}, we focus on a single $K$ point (valley).
The Pauli matrices $\sigma_\alpha$ with $\mbox{\boldmath{$\sigma$}}=(\sigma_1,\sigma_2)$
act in sublattice 
space, and $B(x,y)\hat e_z= {\rm rot}{\bf A}(x,y)$.  The field profiles 
considered in Fig.~\ref{fig1} are independent of the 
longitudinal transport direction $y$
and constant within each of the three regions,
\begin{equation} \label{field}
B(x) = \left \{    \begin{array}{ll}
B,  & x<-d,  \\
B',  & |x|<d,  \\
\sigma B, & x>d,  
\end{array}         \right .
\end{equation}
where $\sigma=\pm 1$ gives
the relative sign of the magnetic field on the two sides $|x|>d$.
We mention in passing that we have also studied the
power-law form $B(x)\propto x^m$ (with
$m=1,2,3$) to make sure that the steps in Eq.~\eqref{field}
do not cause unphysical artefacts.  Indeed
the same qualitative features as reported below for
the profile \eqref{field} were found from such calculations,
which can also benefit from the semi-classical approximation.
A convenient gauge for the vector potential, ${\bf A}=A(x) \hat e_y$ 
with $B(x)=\partial_x A(x)$, is (for $\sigma=+1$)
\begin{equation} \label{vecpot}
A(x) = \left \{    \begin{array}{ll}
Bx+ (B-B')d,  & x<-d,  \\
B'x,  & |x|<d,  \\
Bx- (B-B')d,  & x>d. \end{array} \right .
\end{equation}
Due to translation invariance in the $y$-direction, we can parametrize
solutions $\Psi(x,y)=\psi(x) e^{ik y}$
by the conserved longitudinal momentum $k$.
>From Eq.~\eqref{h0}, for the spinor component $u$ 
in $\psi(x)=(u,v)^T$, we obtain 
 \begin{equation} \label{ueq}
\left[ \partial^2_x - \frac{e}{c}B(x) - \left( k +
\frac{e}{c}A(x) \right)^2 + \epsilon^2 \right] u =0 .
\end{equation} 
For $\epsilon\neq 0$, $ v= \frac{1}{i\epsilon} [ \partial_x - 
k-\frac{e}{c} A(x) ] u$ then gives the other component.
To obtain the bandstructure, we first determine the
general solution in each of the three regions separately.  
Matching conditions follow from the continuity of the wavefunction
at $x=\mp d$ and will be shown to give an energy quantization condition.

\begin{figure}
\includegraphics[width=0.45\textwidth]{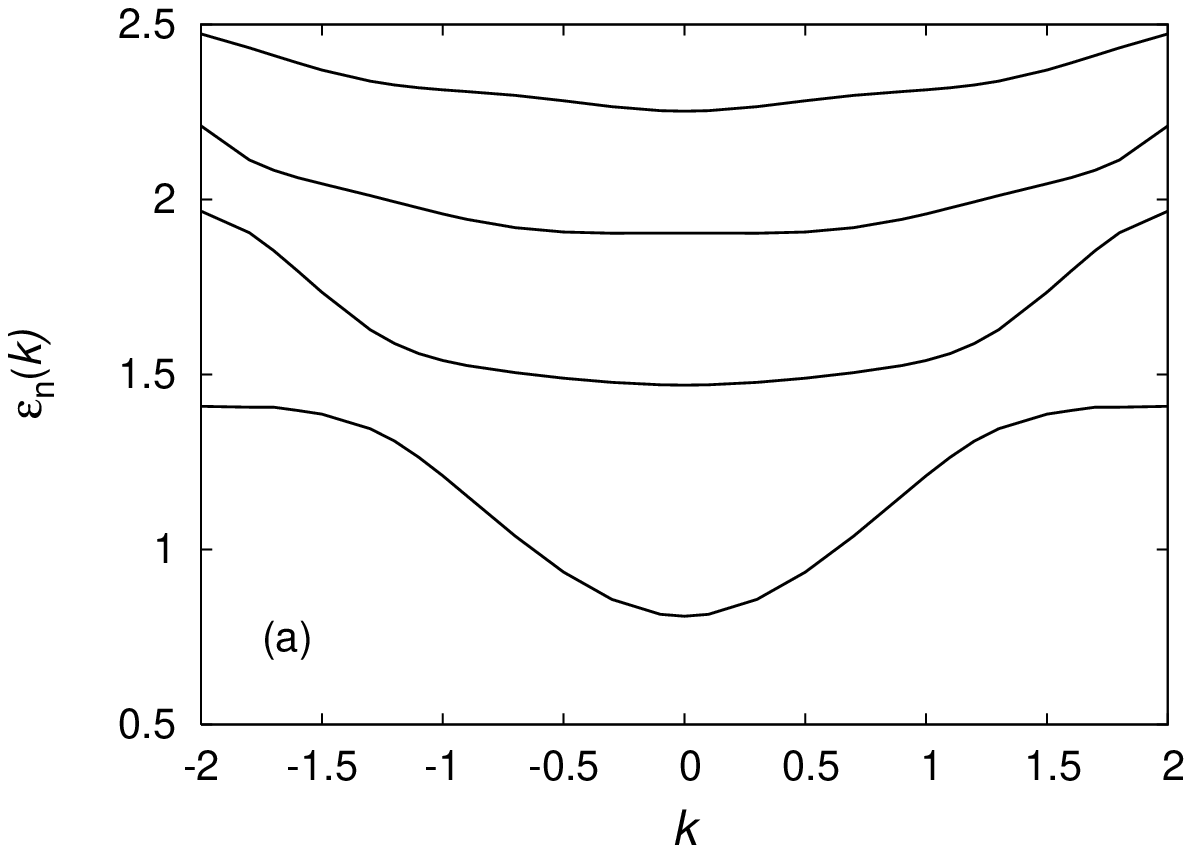}
\includegraphics[width=0.45\textwidth]{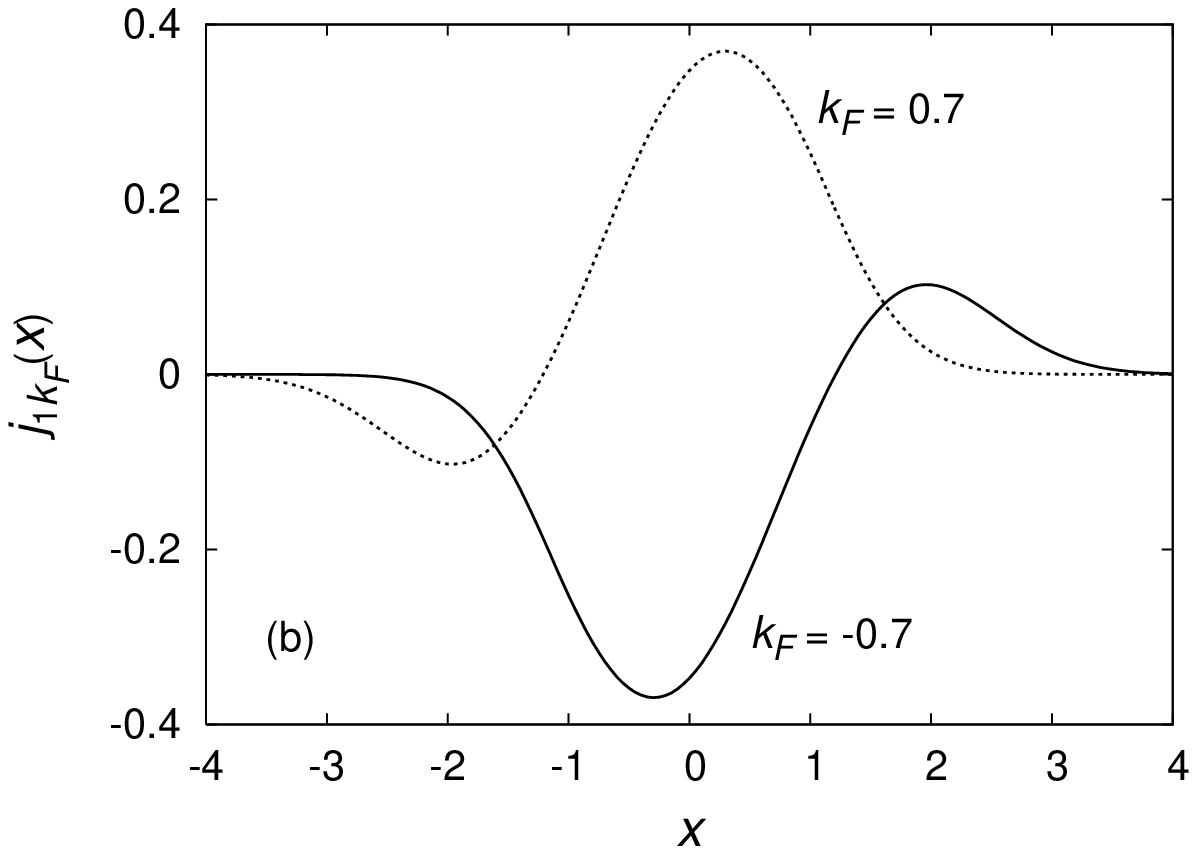}
\caption{ (a) Spectrum of the magnetic waveguide with 
$\sigma=+1$, $d=l_B$ and $B'=0$. 
Energies (momenta) are given in units of $v_F/l_B$ ($l_B^{-1}$).
Only the few lowest electron-like ($\epsilon_n(k)>0$) states are shown.
(b) Current profile $j_{1k_F}(x)$ in units of $v_F/l_B$, see
Eq.~\eqref{curr}, with $x$ in units of $l_B$.  The plot is 
for $n=1$ and $\epsilon l_B=1$, leading to  $k_F l_B\simeq \pm 0.7$.
The two counter-propagating states are centered near the middle
of the waveguide.
 \label{fig2} }
\end{figure}

For $x<-d$, the constant magnetic field $B$ implies
the lengthscale $l_B= \sqrt{c/e|B|}$. We may then explicitly
solve Eq.~\eqref{ueq} in terms of parabolic cylinder functions $D_p(q)$
 \cite{prl}.  With the auxiliary variables  
\begin{equation}\label{aux}
q=\sqrt{2}[(x+d)/l_B + {\rm sgn}(B) k l_B], \quad
p= (\epsilon l_B)^2/2 - 1,
\end{equation}
and complex coefficients $a_\pm$, the solution reads
\begin{eqnarray}
\psi_{B>0}(x) &=&  \sum_\pm a_\pm \left( \begin{array}{l}
D_p(\pm q) \\ \mp \frac{\sqrt{2}}{i\epsilon l_B} D_{p+1}(\pm q)
\end{array} \right), \\
\psi_{B<0}(x) &=& \sum_\pm a_\pm \left( \begin{array}{l}
D_{p+1}(\pm q) \\ \pm \frac{\sqrt{2}}{i\epsilon l_B} (p+1) D_{p} (\pm q)
\end{array} \right).
\end{eqnarray}
Similarly, the eigenfunction for $x>d$ can be expressed with
coefficients $c_\pm$, and replacing $d\to -d$ in Eq.~\eqref{aux}. 
 Finally, for
$B'\neq 0$, the region $|x|<d$ again admits such a representation
with coefficients $b_\pm$ and $d\to 0$ in Eq.~\eqref{aux}.
For $B'=0$, a plane-wave solution applies instead,
\begin{equation}
\psi (x) = \sum_\pm b_\pm \left( \begin{array}{l}
1 \\ \frac{\pm k_\perp+ik}{\epsilon} \end{array} \right) e^{\pm
ik_\perp (x+d)},
\end{equation}
where $k_\perp=\sqrt{\epsilon^2-k^2}$. For $|\epsilon|<|k|$, the 
square root is taken as $k_\perp=i\sqrt{|\epsilon^2-k^2|}$.
Without loss of generality we now put $B>0$.
Normalizability then implies $a_+=c_{-\sigma}=0$ and 
we are left with four complex coefficients, one
of which is fixed by the normalization condition.
The two matching conditions (at $x=\mp d$) for the 2-spinor $\psi(x)$ 
then give 4 equations for 3 unknowns, which generates the sought
condition for the energy bands $\epsilon_n(k)$.

For the {\sl symmetric} setup 
$\sigma=+1$ with $B'=0$, some algebra
yields the  energy quantization condition
\begin{eqnarray}\label{en}
&& w^{-1} (u_2v_1-z^2 u_1 v_2) + w (z^2 u_2 v_1- u_1 v_2)
\\ \nonumber && + (z^2-1) (u_1 u_2-v_1 v_2) = 0,
\end{eqnarray}
which for given $k$ generates an equation for
$\epsilon$ since $k_\perp=k_\perp(\epsilon,k)$. 
Here we used the notation
\begin{eqnarray}\label{udef}
u_{1,2}&=&D_p(\mp\sqrt{2} k l_B),\\ \nonumber
v_{1,2}&=& \pm \frac{\sqrt{2}}{i|\epsilon| l_B} D_{p+1}(\mp \sqrt{2}k l_B),
\\ \nonumber
w&=&(k_\perp+ik)/|\epsilon|,\quad z=e^{2ik_\perp d}.
\end{eqnarray}
Equation \eqref{en} must then be solved numerically, 
and leads to the energy bands $\epsilon_n(k)$ shown
in Fig.~\ref{fig2}(a).  
For large $|k|$, the eigenvalues approach the well-known
relativistic Landau
levels at $\epsilon l_B ={\rm sgn}(n)\sqrt{2|n|}$ \cite{review},
including a zero-energy solution  (not shown in Fig.~\ref{fig2}).

\begin{figure}
\includegraphics[width=0.45\textwidth]{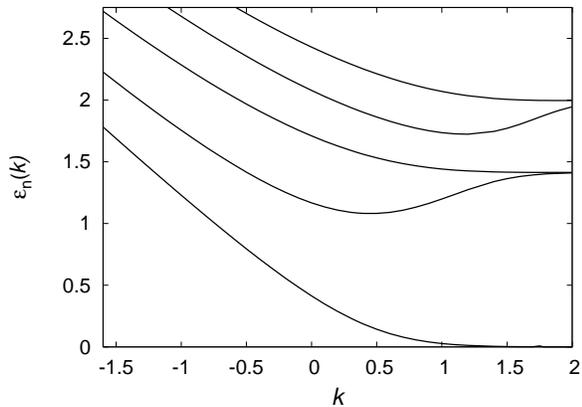}
\caption{ Same as Fig.~\ref{fig2}(a) but for $\sigma=-1$,
cf.\ Fig.~\ref{fig1}(b).
\label{fig3} }
\end{figure}

To illuminate the current-carrying states, we plot
in Fig.~\ref{fig2}(b) the transverse profile of the particle current,
\begin{equation} \label{curr}
j_{nk}(x)= v_F \left(\psi_{nk}^*(x)\right)^T\sigma_2\psi_{nk}(x),
\end{equation} 
where $\psi_{nk}(x)$ is the transverse eigenspinor to energy $\epsilon_n(k)$.
Generalizing the standard argument, see Appendix E in Ref.~\cite{ashcroft},
to the case of Dirac-Weyl quasiparticles, one can show that
\begin{equation}\label{velo}
v_{n}(k) \equiv \int dx j_{nk}(x) = \partial_k \epsilon_n(k).
\end{equation}
We stress that Eq.~\eqref{velo} is a nontrivial result for Dirac
fermions. It holds for any magnetic field profile with $B(x,y)=B(x)$.
This fact leads to the usual cancellation of carrier velocity $v_n(k)$ and 
density of states $(2\pi|\partial_k\epsilon_n(k)|)^{-1}$, 
and thus the two-terminal conductance will be $4e^2/h$
(assuming perfect contacts to reservoirs).
However, as seen in Fig.~\ref{fig2}(b), 
right- and left-moving states occupy the {\sl same}\ spatial region 
and are therefore susceptible
to backscattering perturbations, e.g.\ due to  impurities,
charge inhomogeneities, or fluctuations in the magnetic field.
In practice, quantized conductance is thus
not expected for a waveguide with $B'=0$.

\begin{figure}
\includegraphics[width=0.45\textwidth]{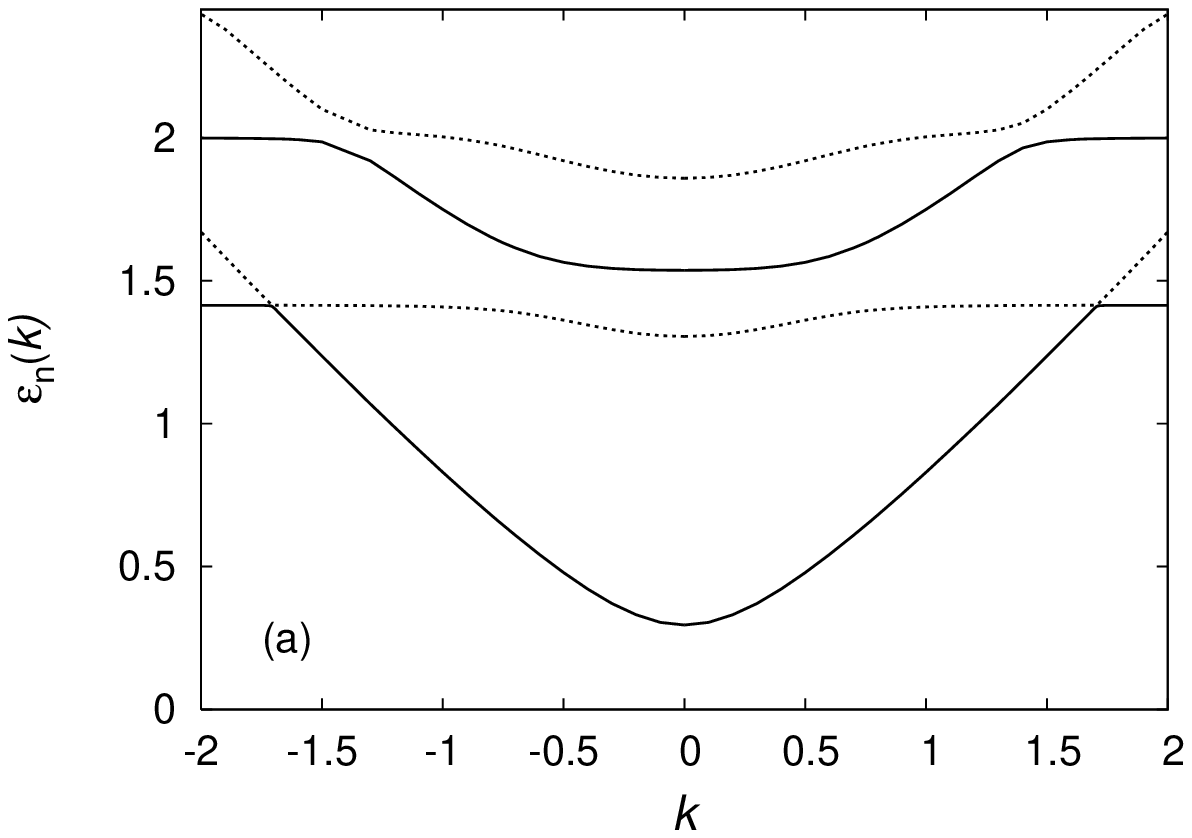}
\includegraphics[width=0.45\textwidth]{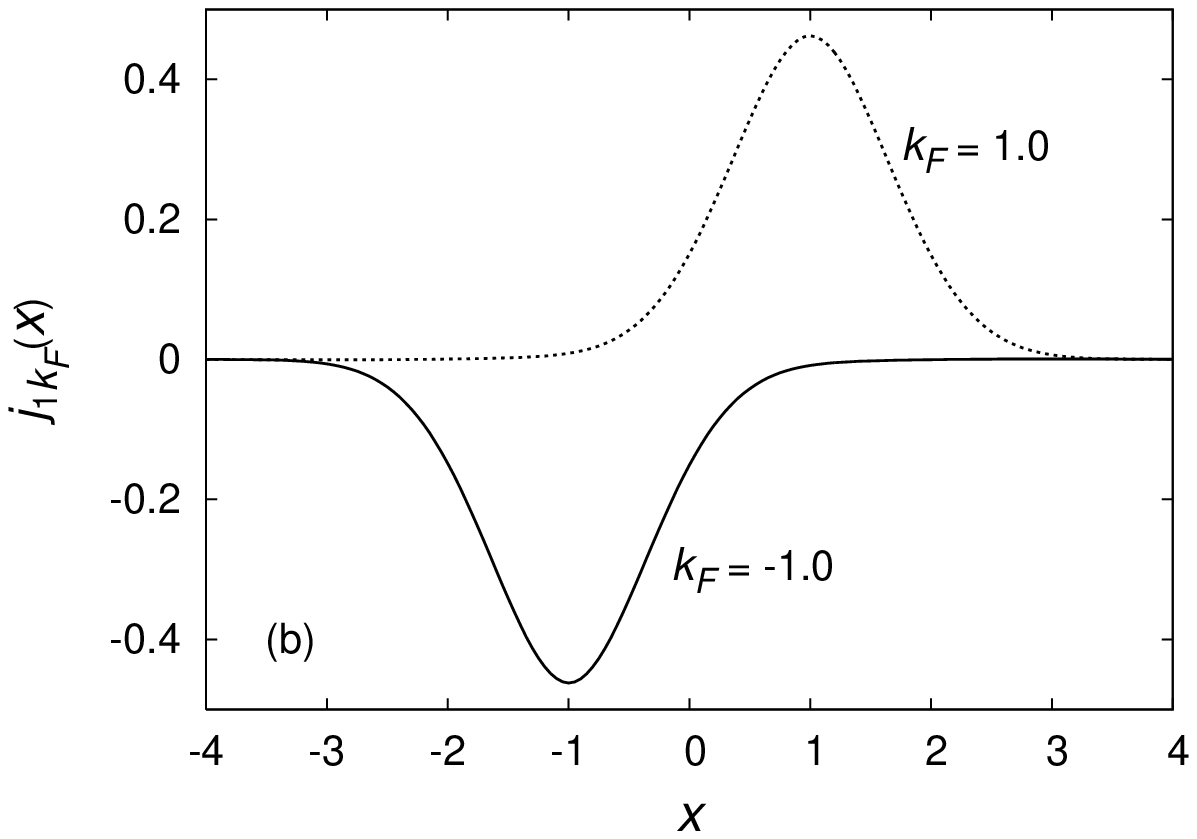}
\caption{ (a) Same as Fig.~\ref{fig2}(a) but for $B'=-B$,
 cf.\ Fig.~\ref{fig1}(a). The lower pair of
$\epsilon_n(k)$ curves has 
an avoided level crossing (not visible on this scale).
The current profile (b) at $\epsilon l_B=0.83$ (corresponding to
$k_F l_B \simeq \pm 1$) shows that the two counter-propagating snake
states are spatially separated already for $d=l_B$.
 \label{fig4} }
\end{figure}

Next we consider the {\sl asymmetric} case with $\sigma=-1$ but still $B'=0$, 
see Fig.~\ref{fig1}(b).  From the 
analogy to Schr\"odinger fermions, one expects to find 
special uni-directional {\sl snake states}  \cite{snake}.
On a semi-classical level, the uni-directionality can be 
understood by noting that cyclotron orbits have a different
winding sense for $x<-d$ and $x>d$. 
The propagating snake state follows by combining half an
orbit from each side and a linear trajectory in the central
region.  The energy quantization condition
takes again the form \eqref{en} after replacing
$u_2=D_{p+1}(-\sqrt{2}k l_B)$
and $v_2= \frac{\sqrt{2}}{i|\epsilon|l_B} (p+1)D_{p}(-\sqrt{2} kl_B)$
in Eq.~\eqref{udef}.  
Numerical solution yields the spectrum depicted in Fig.~\ref{fig3}.
First, we notice a strong asymmetry in the energy bands $\epsilon_n(k)$,
just as in the Schr\"odinger case \cite{snake}.
For $k<0$ a linear dispersion relation is observed, 
corresponding to snake states propagating with $n$-independent 
velocity $|v_n|=v_F$ at sufficiently negative $k$, see Eq.~\eqref{velo}.
The equality of snake velocity and Fermi velocity for $|kd|\gg 1$ also
follows from a simple semi-classical estimate.  Second, the levels merge
pairwise at large positive $k$ to form the relativistic Landau levels,
except for the lowest band in Fig.~\ref{fig3} which merges with the highest
negative-$\epsilon$ band (not shown) to approach the 
zero-energy Landau level.  This is a new feature encountered only for
Dirac fermions
and makes this state easily identifiable for weakly doped graphene.
However,  it is important to stress that for any finite $k$,
there can be {\sl no}\ true
zero-energy state for magnetic field configurations with $\sigma=-1$.
This can be proven on general grounds as a consequence
of the Aharonov-Casher theorem, which in turn follows as a special
limit of the celebrated index theorem \cite{math}.

Interestingly, there is another peculiar subtlety for this magnetic
field profile.  This is seen by computing the equilibrium average
of the current using Eqs.~\eqref{curr} and \eqref{velo},
which predicts a nonzero result.  In fact, the
equilibrium current formally
diverges and is only limited by the bandwidth of the model.
To interpret this non-sensical result we note that in the 
absence of boundaries,  the snake state  propagates in just one
direction and thus produces an {\sl unbalanced current flow}.
The conundrum is resolved when including boundary contributions to 
the current, which are inevitably present in any real sample.
In fact, the dispersion relation in Fig.~\ref{fig3}  ultimately
bends upwards for $k\to \infty$ in the presence of a boundary 
located at $x_b\gg d$. 
The counter-propagating edge state at this boundary will then
balance the total current \cite{snake}.  We have explicitly checked
that this scenario holds true for the case of a zig-zag edge,
where a simple boundary condition on the spinor at $x=x_b$
can be used \cite{ribbonthe}. 
In analogy to quantum Hall edge states \cite{chang}, however,
it should be possible
to experimentally probe the {\sl locally} unbalanced current carried
by the snake state using time-resolved transport measurements \cite{zhitenev}
or scanning tunneling spectroscopy.

We now go back to the {\sl symmetric} setup $\sigma=+1$ but take $B'<0$.
Such a field configuration can be generated by depositing two 
ferromagnetic layers on top of a graphene sheet covered
by a thin insulating layer \cite{expbar,private}. 
In that case one finds two counter-propagating snake states,
and no boundary contributions are required to get zero
total current in equilibrium.  While for $B'=0$,
no snake states exist, they do appear once $B'<0$. 
By generalizing Eq.~\eqref{en}, numerical solution of
the corresponding energy quantization condition leads
to the results in Fig.~\ref{fig4}(a).  
Qualitatively, the spectrum consists of snake states (with approximately
linear dispersion) and Landau level states (dispersionless), with
avoided level crossings between successive eigenenergies $\epsilon_n(k)$.
If the Fermi level intersects only the lowest band shown in Fig.~\ref{fig4}(a),
the quantized conductance $4e^2/h$ follows directly from the Kubo formula.
The current-carrying states at $\pm k_F$ 
are counter-propagating snake states which
are spatially separated and centered near $x=\pm d$, see Fig.~\ref{fig4}(b).
Due to this spatial separation,  weak disorder effects or irregularities
in the magnetic field will not
be able to induce backscattering processes between these states
as long as $d\agt l_B$.
In particular, snake states behave identically for both $K$ valleys,
and thus even inter-valley scattering processes
are  not expected to mix  counter-propagating states.
The conductance quantization in such a setup
should therefore be observable and very precise.

To conclude, we have analyzed the properties of electron waveguides
in graphene, produced by suitable inhomogeneous magnetic field profiles.
Under the setup in Fig.~\ref{fig1}(a) with $B'<0$, we predict robust and highly
accurate conductance quantization in units of $4e^2/h$.
This system is clearly of interest in the context of interacting
1D quantum wire physics, as the electron-electron interaction can lead
to qualitatively new features.  We hope that our work motivates
experimental and further theoretical studies. 

We thank A. Altland, L. Erd\"os, T. Heinzel and J. Smet for discussions.
T.~K.~G.~is supported by the A.~v.~Humboldt foundation.
R.~E.~is supported by the DFG (SFB Transregio 12), and by the ESF
network INSTANS.  

{\sl Note added:} During the preparation of this manuscript, 
a preprint appeared \cite{lambert} where some of our results 
for $\sigma=-1$ were also reported.

\end{document}